\documentclass[10pt,conference]{IEEEtran}
\IEEEoverridecommandlockouts
\def\BibTeX{{\rm B\kern-.05em{\sc i\kern-.025em b}\kern-.08em
    T\kern-.1667em\lower.7ex\hbox{E}\kern-.125emX}}

\usepackage{xspace}
\usepackage{stmaryrd}
\usepackage{comment}
\usepackage{graphicx}
\usepackage{amssymb}
\usepackage{amsfonts}
\usepackage{booktabs}
\usepackage{caption}
\usepackage{tikz}
\PassOptionsToPackage{dvipsnames}{xcolor}
\usepackage{colortbl}

\usepackage{listings}
\usepackage{multirow}

\usepackage{syntax} 
\usepackage{amsmath}
\usepackage{amsthm} 
\usepackage{array}
\usepackage{proof}

\usepackage{hyperref}
\usepackage[capitalise,nameinlink]{cleveref}
\let\autoref\cref

\usepackage{xparse}
\usepackage{algorithm}
\usepackage{algpseudocode}
\usepackage{threeparttable}
\usepackage{enumitem}
\usepackage{tablefootnote}
\usepackage[normalem]{ulem}
\usepackage{minted}
\usepackage{csquotes}
\usepackage[most]{tcolorbox}

\usepackage{stfloats}

\newtheorem{example}{Example}

\begin{document}

\newcommand\yuan[1]{{\color{blue}{\textbf{Yuan:}{\em#1}}}}
\newcommand\yu[1]{{\color{blue}{\textbf{Yu:}{\em#1}}}}
\newcommand\tamjid[1]{{\color{blue}{\textbf{Tamjid:}{\em#1}}}}

\newcommand{\toolname}{{\textsl{AuthFix}}\xspace}

\definecolor{code-color}{rgb}{0.8, 0.47, 0.196}

\newcommand{\hole}{\square\xspace}
\newcommand{\llmquery}{{Q}\xspace}
\newcommand{\net}{\ensuremath{\mathcal{N}}\xspace}
\newcommand{\graph}{\ensuremath{\mathcal{G}}\xspace}

\long\def \ignoreme#1{} 

\newcommand{\declStmt}[2]{{#1}\,{#2}}
\newcommand{\letStmt}[2]{\texttt{let}\,\,{#1} = {#2}}
\newcommand{\assumesStmt}[1]{\texttt{assume}\,\,{#1}}
\newcommand{\ensuresStmt}[1]{\texttt{assert}\,\,{#1}}
\newcommand{\spec}{\ensuremath{\varphi}}
\newcommand{\pred}{\ensuremath{\phi}}

\newcommand{\ie}{i.e.,\xspace}
\newcommand{\Ie}{I.e.,\xspace}
\newcommand{\eg}{e.g.,\xspace}
\newcommand{\Eg}{E.g.,\xspace}
\newcommand{\etAl}{et al.\xspace}

\newcommand{\calA}{\ensuremath{\mathcal{A}}\xspace}
\newcommand{\calB}{\ensuremath{\mathcal{B}}\xspace}
\newcommand{\calC}{\ensuremath{\mathcal{C}}\xspace}
\newcommand{\calD}{\ensuremath{\mathcal{D}}\xspace}
\newcommand{\calE}{\ensuremath{\mathcal{E}}\xspace}
\newcommand{\calF}{\ensuremath{\mathcal{F}}\xspace}
\newcommand{\calH}{\ensuremath{\mathcal{H}}\xspace}
\newcommand{\calI}{\ensuremath{\mathcal{I}}\xspace}
\newcommand{\calK}{\ensuremath{\mathcal{K}}\xspace}
\newcommand{\calL}{\ensuremath{\mathcal{L}}\xspace}
\newcommand{\calM}{\ensuremath{\mathcal{M}}\xspace}
\newcommand\calP{\ensuremath{\mathcal{P}}\xspace}
\newcommand\calQ{\ensuremath{\mathcal{Q}}\xspace}
\newcommand\calR{\ensuremath{\mathcal{R}}\xspace}
\newcommand\calS{\ensuremath{\mathcal{S}}\xspace}
\newcommand\calT{\ensuremath{\mathcal{T}}\xspace}
\newcommand{\calU}{\ensuremath{\mathcal{U}}\xspace}
\newcommand\calV{\ensuremath{\mathcal{V}}\xspace}
\newcommand\calX{\ensuremath{\mathcal{X}}\xspace}
\newcommand\calY{\ensuremath{\mathcal{Y}}\xspace}
\newcommand{\calZ}{\ensuremath{\mathcal{Z}}\xspace}

\newcommand{\den}[1]{\llbracket#1\rrbracket}
\newcommand{\concat}{\rightarrow}
\newcommand{\negate}{!}
\newcommand{\disjunct}{+}

\newcommand \mycirc[1][0.7ex]{\tikz\draw (0,0) circle (#1);} 
\newcommand{\always}{\ensuremath{\square}}
\newcommand{\ltlnext}{\ensuremath{\mycirc}}
\newcommand{\eventually}{\ensuremath{\lozenge}}
\newcommand{\until}{\ensuremath{\mathcal{U}}}

\lstset{escapeinside={(*}{*)},mathescape}
\lstset{basicstyle=\footnotesize\ttfamily,breaklines=true,numbers=left,stepnumber=1}
\lstset{frame=bottomline}

\newcommand \emptycirc[1][1ex]{\tikz\draw (0,0) circle (#1);} 
\newcommand \halfcirc[1][1ex]{%
  \begin{tikzpicture}
  \draw[fill] (0,0)-- (90:#1) arc (90:270:#1) -- cycle ;
  \draw (0,0) circle (#1);
  \end{tikzpicture}}
\newcommand \fullcirc[1][1ex]{\tikz\fill (0,0) circle (#1);}

\newtcolorbox{rqbox}[3][]
{
  colframe = #2!25,
  colback  = #2!10,
  coltitle = #2!20!black,  
  #1,
}

\newtcolorbox{specbox}[3][]
{
  colframe = #2!25,
  colback  = #2!10,
  coltitle = #2!20!black,  
  title    = {#3},
  #1,
}

\newcommand{\yanju}[1]{{{\color{orange}{{\xspace(Yanju: #1)\xspace}}}}}
\newcommand{\tbd}[1]{{{\color{red}{{\xspace #1\xspace}}}}}

\def\sectionautorefname{Section}
\def\subsectionautorefname{Section}
\def\algorithmautorefname{Algorithm}
\def\figureautorefname{Figure}
\def\tableautorefname{Table}
\def\equationautorefname{Equation}

\newcommand{\reducedstrut}{\vrule width 0pt height .1\ht\strutbox depth .1\dp\strutbox\relax}
\newcommand{\icode}[1]{%
  \begingroup
  \setlength{\fboxsep}{0pt}%
  \colorbox{black!10}{\reducedstrut${\texttt{\small #1}}$\/}%
  \endgroup
}


\title{Automated Repair of OpenID Connect Programs (Extended Version)}

\author{
\IEEEauthorblockN{Tamjid Al Rahat\IEEEauthorrefmark{1}, Yanju Chen\IEEEauthorrefmark{2}, Yu Feng\IEEEauthorrefmark{3}, Yuan Tian\IEEEauthorrefmark{1}} \\
\IEEEauthorblockA{\IEEEauthorrefmark{1}\textit{University of California, Los Angeles}, tamjid@ucla.edu, yuant@ucla.edu} 
\IEEEauthorblockA{\IEEEauthorrefmark{2}\textit{University of California, San Diego}, yanju@ucsd.edu} 
\IEEEauthorblockA{\IEEEauthorrefmark{3}\textit{University of California, Santa Barbara}, yufeng@cs.ucsb.edu}
}










\maketitle

\begin{abstract}
OpenID Connect has revolutionized online authentication based on single sign-on (SSO) by providing a secure and convenient method for accessing multiple services with a single set of credentials. Despite its widespread adoption, critical security bugs in OpenID Connect have resulted in significant financial losses and security breaches, highlighting the need for robust mitigation strategies. Automated program repair presents a promising solution for generating candidate patches for OpenID implementations. However, challenges such as domain-specific complexities and the necessity for precise fault localization and patch verification must be addressed. We propose \toolname, a counterexample-guided repair engine leveraging LLMs for automated OpenID bug fixing. \toolname integrates three key components: fault localization, patch synthesis, and patch verification. By employing a novel Petri-net-based model checker, \toolname ensures the correctness of patches by effectively modeling interactions. Our evaluation on a dataset of OpenID bugs demonstrates that \toolname successfully generated correct patches for 17 out of 23 bugs (74\%), with a high proportion of patches semantically equivalent to developer-written fixes. 
\end{abstract}

\section{Introduction}\label{sec:intro} 

Single-Sign-On (SSO) protocols like OpenID Connect are a cornerstone of modern online authentication, supporting billions of accounts and millions of applications across web and mobile platforms~\cite{openid-certification}. The OIDC ecosystem is broad and vibrant: widely used open-source projects such as ory/hydra~\cite{hydra}, dexidp/dex~\cite{dex}, and oauth2-proxy/oauth2-proxy~\cite{oauth2-proxy} form critical infrastructure, while major technology providers maintain official SDKs for developers~\cite{googleapis,cpprestsdk}. This level of adoption makes OIDC both indispensable and high impact, offering seamless login experiences for users, helping organizations manage access and compliance, and mitigating password-related security risks.

Due to its complexity, critical security bugs~\cite{oauthlint-ase19,ccs22} in OpenID Connect can lead to severe consequences, including financial losses and widespread breaches~\cite{attack-oidc,attack-oidc2,news-facebook}. In 2022, a high-severity flaw (with a CVSS score of 8.7) was reported in Google’s authentication flow~\cite{google-bug}, and in 2023, an OpenID-specific bug~\cite{microsoft-bug} in Microsoft’s Azure authentication allowed attackers to forge tokens and compromise over two dozen organizations. These incidents illustrate the difficulty of timely repair, as even well-scoped protocol-level issues can remain unresolved for years. For example, the race condition issue~\cite{issue618} in OAuthLib has been open for {\em over six years}. Therefore, it is crucial to develop approaches that automate patch generation with formal assurance, ensuring vulnerabilities are fixed promptly and reliably.

A natural way to automate the above process is to leverage \emph{program repair}~\cite{le2011genprog,kim2013automatic,xuan2016nopol,xiong2017precise} for generating candidate patches of OpenID implementations. However, this approach faces several \emph{challenges}. First, OpenID is a special domain with few studies in the verification and repair community, which makes it difficult to leverage prior heuristics and data-driven approaches for both fault localization and patch synthesis. Second, the OpenID domain is extremely complex in terms of specification and implementation, which goes beyond the capabilities of existing patch synthesis. For instance, OpenID specification involves hundreds of pages of documentation, making it difficult for developers to conform to all the logical requirements when fixing OpenID bugs. The implementation typically uses many complex APIs (\eg cryptographic APIs) from external libraries. Finally, an incorrect fix can compromise the system's security. For instance, unsigned ID tokens are only supported in OpenID for low-power devices (\eg IoT devices). Accepting unsigned tokens for regular clients like web applications may seem like a convenient shortcut for developers, but it can leave the system exposed to severe attacks like authentication bypass. Therefore, for each candidate patch, OpenID developers need oracles to verify correctness of the patch. While the OpenID Foundation~\cite{openid-certification} offers a certification process, it only evaluates the correctness of protocol endpoints, neglecting the actual implementations. This limitation leaves a critical gap in ensuring the overall functionality and reliability of the repaired programs. Therefore, it is of paramount importance to automate the repair of OpenID bugs while ensuring the correctness of the patches. 

\noindent{\bf Our approach}.\quad To address the above-mentioned challenges, we propose \toolname, a counterexample-guided repair engine powered by large language models (LLMs)~\cite{wu2023large,kang2023preliminary}. Specifically, \toolname takes three inputs: a buggy OpenID program, a specification, and a domain-specific language (DSL), and then returns a fixed version that is guaranteed to conform to the specification. Internally, \toolname is composed of three major components: \emph{fault localization}, \emph{patch synthesis}, and \emph{patch verification}. To address the data sparsity problem due to our specialized domain (i.e., OpenID Connect), both fault localization and patch generation are guided by an LLM (e.g., ChatGPT). Motivated by prior work in component-based synthesis~\cite{neo,sypet}, to address the large space during patch synthesis, \toolname carefully decomposes the problem into two separate phases, namely, \emph{sketch generation} and \emph{sketch enumeration} where invalid partial patches are pruned early. Finally, to ensure the correctness of the candidate patch and the efficiency of the verification, \toolname designs a novel Petri net based model checker, which effectively models the complex interactions among multiple components of an OpenID program. Here, the output of the verification algorithm will be used as the feedback to the \emph{patch synthesis} module to prevent making similar mistakes.

We evaluate our proposed approach on a dataset of OpenID bugs that we compiled based on prior works~\cite{oauthlint-ase19, ccs22, ccs24} and other online platforms. We also collect the patches that are manually written by developers (if available) to fix the corresponding bugs in our dataset. Our evaluation shows that \toolname generates correct patches for 17 OpenID bugs out of 23 that were found in popular OpenID libraries. In addition, our manual assessment shows that 71.4\% of the generated patches are semantically equivalent to the manual patches written by OpenID developers.

In summary, this paper makes the following contributions:
\begin{itemize}
    \item We design a counterexample-guided program repair approach augmented with LLMs to fix real-world bugs in OpenID Connect protocols.
    
    \item We utilize a novel Petri net based approach to validate the patches against the standard OpenID specification.
    
    \item We implement the proposed approach in a tool named \toolname\footnote{The artifact is available at \url{https://github.com/tamjidrahat/authfix}.}, and evaluate it over 23 benchmarks of real-world OpenID Connect bugs. Our evaluation shows that \toolname can generate patches for 74\% (17) of bugs that were discovered and reported in previous work.
\end{itemize}

\section{Background}\label{sec:background}
We provide a background of OpenID Connect (OIDC) protocol and explain the security impacts of OIDC bugs. 
\subsection{OpenID Connect}
Among all single sign-on (SSO) protocols, OIDC is the most popular and supported by nearly all major identity service providers, including Google, Microsoft, and Amazon. OIDC is an authentication protocol~\cite{openid-spec} allowing users to authenticate themselves across different web and mobile applications. OIDC utilizes an identity layer added on top of the authorization flows, which allows the clients to authenticate end users using their existing accounts. Precisely, a user authenticates with the OIDC provider's authorization server and receives an ID token, a JWT-formatted string~\cite{jwt} that contains information about the user's identity, such as their name and email address as payload along with a signature component that can be cryptographically verified by the recipient. The ID token is then used to authenticate the user to a web or mobile application, which can then authorize the user to access resources or services. The protocol relies on communication between multiple parties, such as the Relying Party (RP), OIDC Provider (OP), and users. 

\begin{figure}[t]
\centering
  \includegraphics[width=\linewidth]{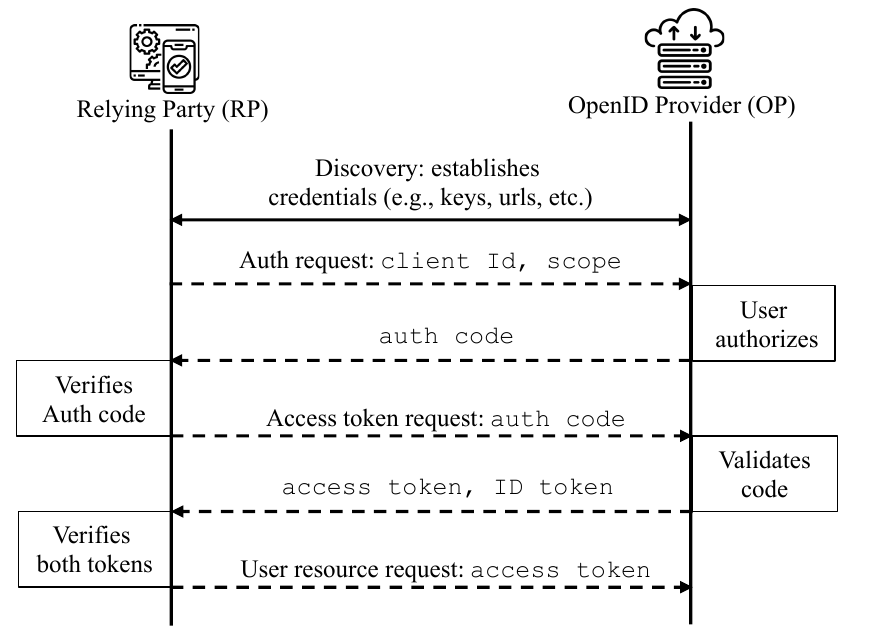}
  \caption{Interactions between the Relying Party (RP)  and OpenID Provider (OP) during the authentication process of \textit{Authorization Code Flow} in OIDC.}
  \label{fig:protocol-flow}
  \vspace{-1.5em}
\end{figure}

OIDC provides support for three authentication flows that can be implemented by the participating parties: (1) Authorization Code Flow, (2) Implicit Flow, and (3) Hybrid Flow. Each flow defines the transactions that occur between multiple parties during the authentication process. \autoref{fig:protocol-flow} illustrates an example of an Authorization Code Flow, where upon establishing the discovery of credentials, RP initiates an authentication request. The OP then issues an authorization code upon successful authorization from end users. RP then verifies the code and exchanges it for an access token along with an ID token. Finally, RP verifies the authenticity of the tokens and exchanges the access token for user resources.

\begin{figure*}[t]
\centering
  \includegraphics[width=\linewidth]{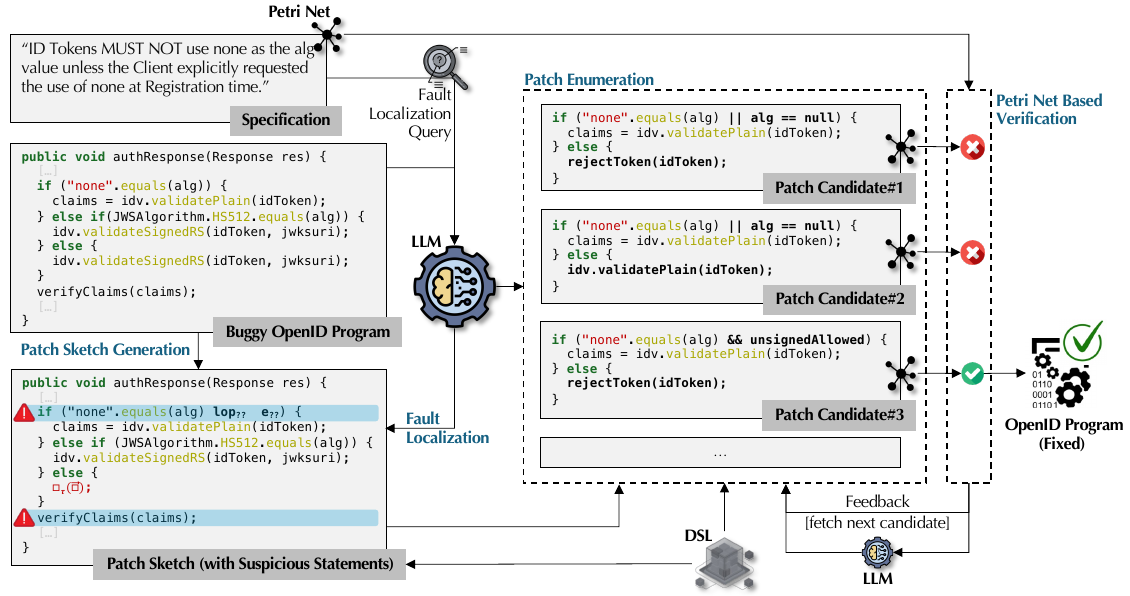}
  \caption{Workflow of \toolname on a concrete example. Here, \toolname first takes a buggy OpenID program and its corresponding English specification as inputs and then generates repair sketches based on the fault localization by an LLM agent. The tool then performs patch enumeration while transforming the repair candidates into their corresponding Petri-net models, which are verified against the Petri-net specification constructed (manually) from the English specification.}
  \label{fig:workflow}
  \vspace{-1.5em}
\end{figure*}

\subsection{Security Impacts of OpenID Bugs}
OIDC is a widely used protocol for authentication and authorization in modern software. The prevalence of logical bugs in the OIDC implementations is due to several factors such as implementation mistakes, lack of adherence to the best practices, and complexities of the specification. Following are some of the most common categories of bugs in OIDC: 
\begin{itemize}[leftmargin=*]
    \item Authorization Code Interception: The authorization code flow is susceptible to interception attacks if the code is processed without proper validation, or if the redirect URI is not adequately protected. Attackers can intercept the authorization code and use it to obtain access tokens.
    
    \item Cross-Site Request Forgery (CSRF): CSRF attacks occur when an attacker tricks a user into unknowingly executing actions on a website on which the user is authenticated. Incorrect use of the ``\textit{state}'' parameter during the protocol execution can lead to unauthorized actions being performed using the authenticated session.

    \item Token Expiry and Revocation: Failure to enforce token expiration policies or properly handle token revocation can lead to unauthorized access. Tokens should be regularly refreshed, and revoked tokens should be invalidated promptly. 

    \item Unauthorized Access to Account: If an attacker successfully injects a malicious token into the OIDC flow, they can gain unauthorized access to protected resources or user accounts. They might forge an access token with elevated privileges or modify an ID token to impersonate a legitimate user.
\end{itemize}

\section{Overview}\label{sec:overview}

Here, we present an overall workflow of \toolname with a concrete example presented in \autoref{fig:workflow}. The example represents bugs inspired by a recently reported vulnerability (CVE-2021-44878) from Pac4j library which provides Java-based security framework for web applications. The library utilizes OIDC's ID token to provide secure authorization and authentication support for a wide variety of platforms. This example demonstrates a flaw in the ID token validation process, potentially allowing attackers to bypass signatures and inject malicious payloads, such as access token, into OpenID entities. Specifically, the library incorrectly handles the algorithm (\textit{alg}) parameter when verifying the signature component of the ID token. Since the \textit{alg} parameter is encoded in the header component and can be altered by the attacker, it can be leveraged to bypass the signature-based authentication. Despite being classified as a high-severity bug, developers took nearly a month to patch it due to the inherent complexity of correctly validating ID token according to the specifications. 

The example includes a set of validity checks, including the validation of the signature and certain OpenID-specific claims (\eg issuer, audience, etc.). Specifically, to perform the signature validation for ID token, this example utilizes the value of \textit{alg} parameter in the header component of the token. Unfortunately, as shown in line 3, the code does not perform a check required by the OpenID specification~\cite{openid-spec}, which states that ``\textit{ID tokens must not use \icode{none} as the \icode{alg} value unless it is explicitly requested by the client during the registration time}.'' Therefore, \icode{validatePlain} method should not be invoked unless the boolean \icode{unsignedTokenAllowed} in configuration is \icode{true}. In addition, the code incorrectly invokes the \icode{validateSignedRS} method (line 8) when the \icode{alg} value does not match any of the expected values (e.g., HS256, RS256.) specified by the protocol. The specification requires rejecting the ID token in such a case. ID token validation is critical for OpenID as it is the cornerstone of authentication in OIDC. Specifically, ID token issued by OP consists of a signed component that allows the RP to verify the authenticity of the information received from OP.

\begin{figure*}[t]
\centering
  \includegraphics[width=0.7\linewidth]{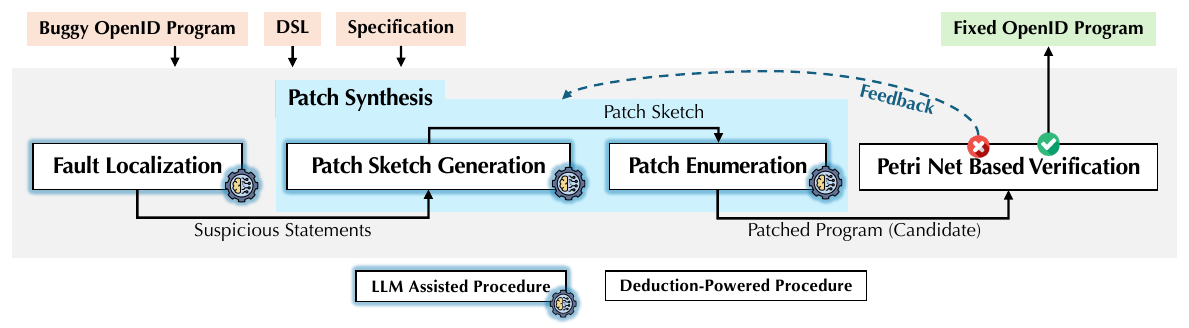}
  \caption{An overview of \toolname.}
  \label{fig:overview}
  \vspace{-1.5em}
\end{figure*}

Without proper validation of ID token, the authentication process of OpenID can be vulnerable to access token injection which would lead to unauthorized access or even user account takeover by attackers. For example, in the context of the above code example, using \icode{none} as \icode{alg} value forces the protocol to {\em skip the signature validation}, attackers can alter and inject an ID token with malicious claims and bypass the signature validation. However, automatically fixing this buggy code is highly non-trivial as it requires a comprehensive understanding of the OpenID specification and exploring a large space of candidate patches that involve multiple locations. Consequently, it takes months for developers to fix such highly sensitive bugs. For instance, a vulnerability (CVE-2021-22573) in the ID token validation of Google's authorization library remained unpatched for nearly four months, leaving both the library and the applications that depended on it exposed to risk for an extended period. 

\noindent{\bf Motivation and our solution}.\quad Repairing OpenID Connect programs requires addressing both the complexity of the specification and the semantic correctness of fixes. OIDC protocols involve concurrent interactions across multiple parties, with correctness conditions spanning sequential steps and synchronization points. Test-based or heuristic repair cannot ensure compliance in this setting, since many bugs arise from protocol-level violations rather than local code errors. A formal model is therefore needed to capture both sequencing and concurrency in authentication flows.

AuthFix builds on a Petri-net–based model of OIDC, which represents protocol states, concurrent transitions, and forbidden behaviors specified in the standard, enabling validation of candidate repairs against the full specification. On this basis, AuthFix follows a counterexample-guided inductive synthesis (CEGIS)~\cite{brahma,sketch} paradigm, where candidate patches are generated, verified against the Petri-net model, and iteratively refined with counterexamples until compliance is achieved. Concretely, the workflow contains the following steps:
\begin{itemize}
    \item {\em Preparation}.\quad The specification of security-sensitive OIDC components (e.g., authorization flow, ID token verification) is first obtained from the official standard~\cite{openid-spec} and then manually transformed into an equivalent Petri-net representation, enabling modeling of the OIDC flow together with its concurrent requirements that must hold during protocol execution. During the repair process, the buggy program and its corresponding specification are embedded into an LLM query to identify suspicious statements.

    \item {\em Sketch Generation}.\quad Based on these statements, a set of schemas is applied to generate repair sketches containing partial expressions represented as holes (denoted by $\hole$). For example, in \autoref{fig:workflow}, line 3 of the buggy program is transformed into the {\em partial} expression \icode{none.equals(alg) $\hole_\sigma$ $\hole$}, where $\hole_\sigma$ denotes a hole for logical operators and $\hole$ denotes a hole for expressions. Similarly, the function call in line 8 is transformed into $\hole_\tau(\vec{\hole})$, which includes a hole for a method $\hole_\tau$ of type $\tau$ and its arguments $\vec{\hole}$.

    \item {\em Patch Synthesis}.\quad Guided by a domain-specific grammar optimized by an LLM agent, AuthFix then generates concrete repair candidates. For instance, the operator hole $\hole_\sigma$ may be instantiated as $\&\&$ or $||$, producing multiple candidates.

    \item {\em Verification}.\quad Each candidate is translated into an equivalent Petri net and validated against the specification-derived Petri net. This process iterates until a valid patch is found or the search space is exhausted.

    \item {\em Counterexample-Guided Refinement}.\quad When verification fails, the counterexample trace provided by the Petri-net checker is fed back into the synthesis loop, guiding the LLM to refine subsequent candidates and converge toward a specification-compliant repair.
\end{itemize}
Together, these stages constitute a CEGIS-style repair framework that ensures specification-compliant fixes, and the following section elaborates on each component in detail.
\section{Methodology}\label{sec:method}

\autoref{fig:overview} presents the overall architecture of \toolname. Given a buggy OpenID program, a specification, and a DSL, \toolname executes three main procedures: fault localization, patch synthesis (including sketch and candidate enumeration), and Petri-net–based model checking, producing a repaired version certified by the verifier. In the first step, suspicious statements are identified through fault localization, from which a \emph{sketch}---a partial program with unfilled holes---is derived. These holes are then completed through best-first enumeration to construct concrete repair candidates. To ensure correctness, \toolname translates each candidate into a Petri net and validates it against the specification using the verification procedure. Verification yields two outcomes: (1) the patched program is accepted if it conforms to the specification; or (2) otherwise, the procedure returns feedback (e.g., counterexamples) that exposes the root cause of inconsistency, prompting patch synthesis to generate new candidates until a verified repair is found or the search times out.

To avoid ad-hoc heuristics that are hard to generalize across benchmarks, \toolname interacts with large language models (LLMs) for prioritizing choices in fault localization and patch synthesis. For example, suspicious statements are labeled by LLMs during the invocation of the fault localization procedure, and patches are ranked within the patch synthesis procedure before they get checked by the verification procedure.

In what follows, we elaborate on different components of \toolname. Because patch verification is one of our major contributions, we defer its detailed discussion to \autoref{sec:verification}.

\subsection{Fault Localization}

To ensure the correctness of the patch, existing fault localization methods~\cite{abreu2007accuracy,abreu2009spectrum,pearson2017evaluating} rely on a comprehensive set of test cases, which are not available and hard to construct for most OpenID implementations. To mitigate this limitation, \toolname utilizes an LLM agent, which contains a \emph{query process} and a \emph{response process} modules, to locate suspicious statements that might cause the bug.

\subsubsection{Construction of LLM Query}

Our fault localization approach begins with a text-formatted query \llmquery to the LLM agent. An LLM query \llmquery is a sequence of text-based components $\llmquery = [q_0, q_1, \dots, q_n]$, where each component $q_i$ is one of the following:
\begin{itemize}
    \item {\em Static query component}. A static query component remains the same for all benchmarks, i.e., $q_i(p) = q_i(p')$ for all benchmark programs $p$ and $p'$. We use static components to provide generic instructions (e.g., output format description) for the LLM agent.

    \item {\em Dynamic query component}. Dynamic query components may differ across the benchmarks, i.e., there may exist benchmark program $p$ and $p'$ for which $q_i(p) \neq q_i(p')$.
\end{itemize}
In particular, a dynamic component is one of the following:
\begin{itemize}
    \item {\em Buggy code}. This component includes the body of the function containing the OpenID bugs.
    
    \item {\em Specification}. This component describes the expected behavior of the buggy code extracted from the standard OpenID Connect specification~\cite{openid-spec}, which is written in plain English. 
\end{itemize}

\subsubsection{Processing of Query Response}

\begin{figure}[t]
\centering
  \includegraphics[width=\linewidth]{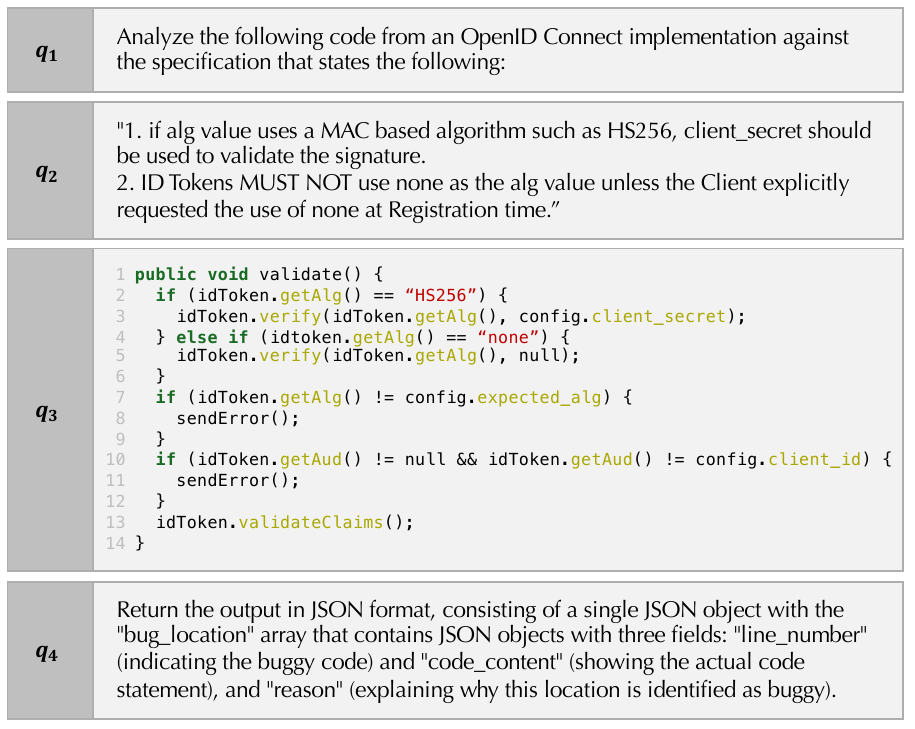}
  \caption{An example of an LLM query for fault localization in a partial OpenID Connect program. This query has four text components \llmquery = [$q_1$, $q_2$, $q_3$, $q_4$], where $q1$ and $q_4$ are static components that provide instructions to the LLM agent and $q_2$ and $q_3$ are dynamic components that provide the specification and buggy program, respectively.}
  \label{fig:llm-query}
  \vspace{-1.5em}
\end{figure}

The LLM queries produce output in JSON format, as specified in the query. \autoref{fig:llm-output} illustrates an example of the output generated for the LLM query in \autoref{fig:llm-query}. The generated output consists of a JSON object containing an array of fault localization results, where each object provides detailed information about the bug, including its specific location (line number), the code content of the corresponding buggy statements, and the reason for the bug. After localizing the bug locations, we generate repair sketches that utilize a series of schemas to replace the buggy statements with partial expressions containing placeholders (denoted by as $\hole$, or {\em holes}) that need to be filled in with concrete repair expressions.

\subsection{Patch Synthesis}

To generate potential repair candidates, we reduce the problem of program repair to program synthesis by transforming the buggy program into the patched version. To avoid state explosion in synthesis, Depending on the type of suspicious statements, we carefully design a set of schemas to generate the sketches. In the following, we elaborate on the details.

\subsubsection{Schemas for Generating Repair Sketch}

\toolname applies the following set of schemas to generate program sketches for repairing based on the identified bug locations in the input program. 

\paragraph{\bf Conditional statement}
Given a buggy conditional statement $e$, \toolname applies the following schemas {\em in the presented order} for sketch generation:
\begin{itemize}
    \item \textbf{(S1)} transforms any arithmetic, relational or logical operator $\sigma$ into an operator hole $\hole_\sigma$ of the same type.  
    
    \item \textbf{(S2)} transforms any variable, constant or field dereference of type $\tau$ into an expression hole $\hole_\tau$ of the same type.

    \item \textbf{(S3)} introduces a new top-level computation for $e$, yielding \icode{$e$ $\hole_\sigma$ $\hole$}, where $\hole_\sigma$ denotes a logical operator and $\hole$ corresponds to an expression of compatible type.
\end{itemize}

\begin{figure}[t]
    \begin{minted}[xleftmargin=5pt,numbersep=1pt,fontsize=\scriptsize,linenos,breaklines]{json}
{"bug_location": [
  {
    "line_number": 5,
    "code_content": "idToken.verify(idToken.getAlg(), null);",
    "reason": "ID tokens MUST NOT use 'none' as the alg value unless explicitly requested."
  },
  {
    "line_number": 7,
    "code_content": "if(idToken.getAlg() != config.expected_alg) {",
    "reason": "Algorithm verification does not handle 'none' algorithm correctly."
  }
]}
    \end{minted}
    \caption{An LLM response for the query described in \autoref{fig:llm-query}.}
    \label{fig:llm-output}
    \vspace{-1.5em}
\end{figure}

\paragraph{\bf Method invocation}
Given a buggy method invocation (\eg $m(\dots)$), \toolname applies the following set of schemas for sketch generation:
\begin{itemize}
    \item \textbf{(S4)} transforms the object reference $o$ of type $\tau$ with an expression hole $\hole_\tau$.
    
    \item \textbf{(S5)} replaces the argument expression $e$ of type $\tau$ with an expression hole $\hole_\tau$.

    \item \textbf{(S6)} transforms the method invocation $m(\vec{a})$ of type $\tau$ into $\hole_\tau(\vec{\hole})$ where $\hole_\tau$ corresponds to methods of the same type and $\vec{\hole}$ corresponds to its arguments.

    \item \textbf{(S7)} adds a {\em conditional guard} $\hole_{\tau={\sf bool}}$ to the method invocation $m(\dots)$, yielding \icode{${\sf if}\ \hole_{\tau={\sf bool}}\ {\sf then}\ m(\dots)$}, which enables additional pre-condition check before invocation.
\end{itemize}

\paragraph{\bf Return statement (S8)}
Given a buggy return statement $e$ of type $\tau$, \toolname transforms it into $\hole_\tau$, an expression hole of the same type.

\subsubsection{Synthesis of Repair Expression}

\toolname leverages a rich set of repair expressions to synthesize the holes in the generated sketches and thereby generate the potential repair candidates. Given a sketch with holes near location $l$, relevant expressions used for repair are determined by: (a) relevant program elements in scope at $l$ and (b) OpenID-specific elements that are available in the context of the underlying authentication flow of the program. 

To synthesize the holes, \toolname extracts all the local variables and literals in scope, fields in the same class and public fields from other classes that are relevant to the buggy class of the program. The relevant classes extracted are based on the classes that are instantiated, whose fields or methods are accessed within the buggy method. 

In addition to the program elements within the scope, \toolname further extracts all the OpenID-specific elements that can be accessed within the context of the underlying OpenID \textit{authentication flow} for the buggy program. For instance, if the buggy program is part of an authentication flow such as implicit flow or hybrid flow of OpenID Connect, we extract the configured values (i.e., values determined during the registration process), constants, and external functions that are globally accessible during the execution of authentication flows. 

\begin{figure}[t]
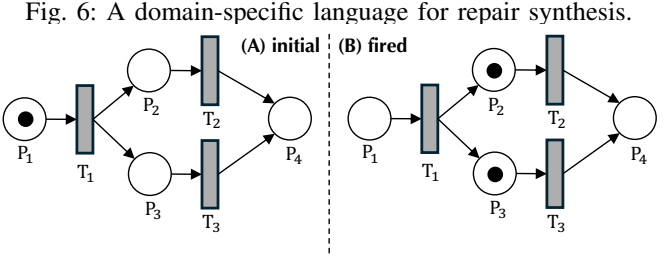

    {\footnotesize
    \[
    \begin{array}{lrll}
        {\sf constant} & {\sf const} & := & \calC \ | \ {\sf null} \ | \ {\sf true} \ | \ {\sf false} \ | \ {\sf id} \ | \ k \\
        {\sf arithmetic\ op} & {\sf aop} & := & + \ | \ - \ | \ \times \ | \ / \ | \ \% \\
        {\sf relational\ op} & {\sf rop} & := & == \ | \ != \ | \ > \ | \ < \ | \ \geq \ | \ \leq \\
        {\sf logical\ op} & {\sf lop} & := & \&\& \ | \ || \\
        {\sf atomic\ expr} & {\sf expr} & := &  {\sf var} \ | \ {\sf const} \ | \ {\sf var}.f\\
        {\sf composite\ expr} & {\sf expr} & := & {\sf expr} \ {\sf op} \ {\sf expr} \ | \ {\sf var}[{\sf expr}]\\
        {\sf arguments} & {\sf args} & := & {\sf args}, {\sf expr} \ | \ {\sf expr}\\
        {\sf invocation} & {\sf m} & := & \mathcal{F}({\sf args}) \ | \ {\sf var}.\mathcal{F}({\sf args})\\
    \end{array}
    \]
    \[
    \begin{array}{lrl}
        \calC \in \textbf{OpenID Constants}\quad\quad \calF \in \textbf{OpenID Functions} 
    \end{array}
    \]
    }
    \caption{A domain-specific language for repair synthesis.}
    \label{fig:expression-syntax}
    \vspace{-1.5em}
\end{figure}

\paragraph{Syntax of repair expressions}
\autoref{fig:expression-syntax} illustrates the grammar of the repair expressions used by \toolname to complete the holes in the sketches. We define the non-deterministic choices of program constructs that include expressions, operators, and method invocation. The atomic expression holes represent the program variables, constants, and field accesses. Constants include both program constants and OpenID-specific constants that are relevant within the scope of the holes. We define arithmetic operators, relational operators, and logical operators to complete the operator holes in the sketches. Moreover, we synthesize composite expression holes by combining expression holes with operator holes. Composite expressions can further be combined together to synthesize more complex expressions. Method invocation holes include OpenID-specific methods along with the list of arguments associated with the methods. 

The grammar of the repair expressions is then used by \toolname to instantiate a program transformation at each location of the holes of the sketches to produce a set of concrete repair candidates. Specifically, \toolname uses an enumeration-based search to synthesize generalizable repair expressions for each type of hole in the sketches. Instead of processing sketches and holes arbitrarily, we process them in the order of sketch generation schemas mentioned above. Specifically, the sketch generated by the schema (S1) is processed before the sketch by (S2). If the search procedure cannot find a solution for the sketch by a schema, it advances to the next. This approach tractably constrains the search space, especially for the bugs with simple fixes (e.g., alternative operator). While synthesizing a hole in a given sketch, if the search fails, \toolname iteratively chooses a new element from the grammar and generates a new repair candidate. Each candidates are then validated against the specification using our model checker.

\paragraph{LLM-driven search space pruning and prioritization} 

Using the grammar in \autoref{fig:expression-syntax} may generate a significantly large number of candidates of repair expressions, especially for composite expressions and method invocation. Since LLM has shown success in understanding the semantics of a given program~\cite{jin2023inferfix}, we utilize the power of LLMs to prioritize the expressions that have a high likelihood of being valid for fixing the underlying bug. To do so, we include the program sketches and submit a query to LLM to suggest candidate expressions for each hole in the program. Therefore, in our iterative search for repair expressions for the holes, we first choose the expressions suggested by the LLM agent before other expressions generated from the grammar. 
\section{Petri-Net–Based Verification}\label{sec:verification}

\begin{figure}[!t]
    \centering
    \includegraphics[width=\linewidth]{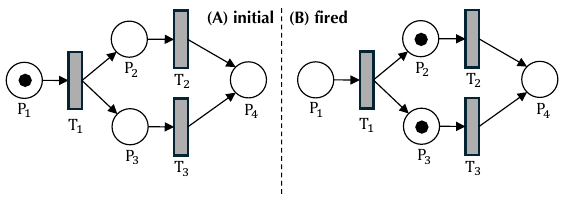}
    \caption{(A) A Petri net with four places (circles) and two transitions (rectangles); (B) The same Petri net after firing $T_1$.}
    \label{fig:petrinet-simple}
    \vspace{-1.5em}
\end{figure}

In this section, we introduce the Petri net based verification framework of \toolname. The goal is to formally check whether a candidate patch conforms to the OpenID specification by modeling both the program and the specification as {\em guarded Petri nets} and analyzing their reachable states. We first review the basics of Petri nets and reachability (\autoref{subsec:petri-nets}), then extend them with guarded transitions (\autoref{subsec:guarded-nets}) and show how to construct nets from both code (\autoref{subsec:code2net}) and specification (\autoref{subsec:spec2net}), followed by the validation procedure (\autoref{subsec:validation}) and the use of counterexamples to refine patches (\autoref{subsec:cex}).

\subsection{Petri Nets and Reachability}\label{subsec:petri-nets}

A Petri net is a bipartite graph $\langle P,T,A,M_0\rangle$:  places $P$ hold \emph{tokens}, transitions $T$ move tokens along arcs $A$, and $M_0$ is the initial marking. A transition fires when every input place has at least one token, after which it consumes those tokens and produces its outputs.

A marking $M \in R(M_0)$ is {\em reachable} from $M_0$ if there exists a transition sequence $\sigma\in T^{*}$ such that $M_0 \xrightarrow{\;\sigma\;} M$, where $R(M_0)$ denotes the {\em reachable set}.

\begin{example}\rm
    \autoref{fig:petrinet-simple}(A) starts with $M_0=(1\,0\,0\,0)$.  Firing $T_1$ yields \autoref{fig:petrinet-simple}(B), where $M_1=(0\,1\,1\,0)$, and then $T_2$ yields $M_2=(0\,0\,0\,1)$. Here, we have $M_1, M_2 \in R(M_0)$.
\end{example}

\begin{figure*}[!t]
    \centering
    \includegraphics[width=\linewidth]{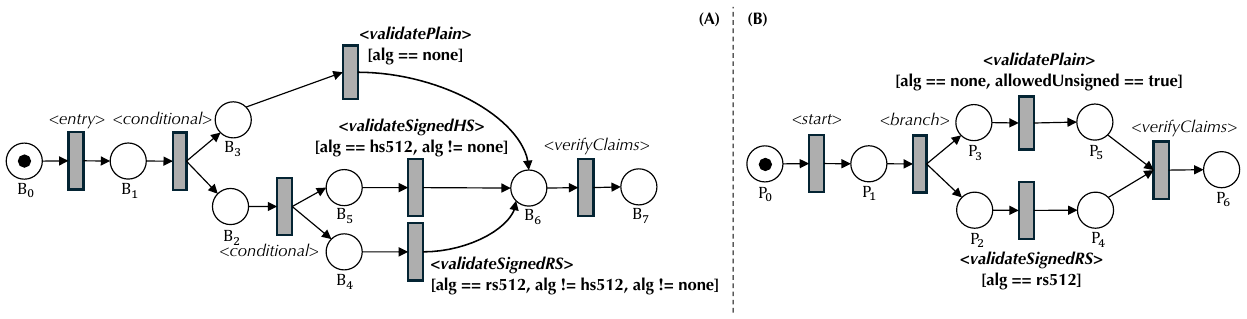}
    \caption{(A) A Petri net example (simplified) for the example in \autoref{fig:workflow}, and (B) Petri net specification provided by users.}
    \label{fig:program-spec-petri}
    \vspace{-1.5em}
\end{figure*}

\subsection{Petri Net with Guarded Transitions}\label{subsec:guarded-nets}

We extend the Petri net described above with guarded conditions, which allows imposing additional constraints on Petri net transitions. Specifically, we extend the transition $T$ to a pair $\langle \epsilon, \gamma \rangle$, where $\epsilon$ denotes a transition event and $\gamma$ denotes the guard condition corresponding to the transition. Therefore, in a given execution environment, a transition $T_i = \langle \epsilon_i, \gamma_i \rangle$ is enabled if and only if event $\epsilon_i$ occurs and the guard condition $\gamma_i$ is satisfied. Guards are defined over the visible variables at the incoming places using binary operators (e.g., $>, >=, ==, \dots, \&\&, ||$, etc.).

\subsection{Construction of Petri Nets from Codes}\label{subsec:code2net}

Given a method’s control-flow graph $\mathsf{CFG} = \langle N, E\rangle$ we build a guarded Petri net $\mathcal N_P = \langle P, T, A, M_0\rangle$ as follows:

\begin{itemize}
    \item {\bf Places $P$.} Create one place $p_n$ for every basic block $n \in N$; a token in $p_n$ denotes that the program counter is currently in that block.

    \item {\bf Transitions $T$.} For each conditional edge $(n_i,n_j) \in E$ add a branch transition $t_{i\rightarrow j}$ guarded by the edge predicate (e.g., \icode{alg==none}). For every call site generate a transition labelled with the callee name; its guard is \texttt{true} unless a value condition is present (e.g., \icode{role.equals("admin")}). Two distinguished transitions $t_{\sf entry}$ and $t_{\sf exit}$ mark the beginning and end of the method so that inter-procedural composition is possible.

    \item {\bf Arcs $A$.} Insert an arc $p_{n_i} \to t_{i\rightarrow j}$ and an arc $t_{i\rightarrow j} \to p_{n_j}$ for every control-flow edge $(n_i,n_j)$; all arc weights are one.

    \item {\bf Initial marking $M_0$.} Place a single token in the entry block’s place. One token suffices because we analyse one dynamic execution of the method at a time; recursion or concurrency is handled at the call-graph level.
\end{itemize}

\autoref{fig:program-spec-petri}(A) shows the resulting net for our running example.  Blocks $B_0 \ldots B_6$ map to places; transitions such as ${<}{\sf validatePlain}{>}$ carry the guard \icode{alg==none}. \autoref{sec:verification} checks and rejects any candidate patch with markings that are not reachable.

\subsection{Construction of Petri Nets from Specification}\label{subsec:spec2net}

\toolname uses a Petri net representation of the OpenID Connect specification which defines the expected behavior during the protocol execution. Specification Petri net is defined as $\net_{\phi}(P_{\phi}, T_{\phi}, A_{\phi}, {M_0}_{\phi})$ where,
\begin{itemize}
    \item $P_{\phi}$ is the set of protocol's execution state;
    \item $T_{\phi}$ is the set of OpenID events that changes the protocol's state. These events can  be constrained via \textit{guard} conditions;
    \item $A_{\phi}$ is the set of arcs that connect the execution states and transitions; and
    \item ${M_0}_{\phi}$ is the initial state of the protocol.
\end{itemize}

To derive $\mathcal{N}_\phi$ from the natural-language description of each OpenID Connect authentication flow (Authorization Code, Implicit, and Hybrid), we follow a systematic mapping procedure:
\begin{itemize}
    \item {\bf Execution states as places.} Each major step in the OIDC protocol (e.g., discovery, authorization request, issuance of tokens, ID token validation, and claim verification) is mapped to a place in $P_\phi$. A token in a place denotes that the execution has reached the corresponding stage of the protocol.

    \item {\bf Events as transitions.} Each message exchange or logical step prescribed by the specification is represented as a transition in $T_\phi$. For example, moving from ``authorization granted'' to ``token exchange'' is modeled as a transition consuming the token from one place and producing it in the next.

    \item {\bf Guards from specification constraints.} Conditional requirements in the specification (e.g., ``ID tokens must not use none as the alg value unless ...''~\cite{openid-spec}) are encoded as guard conditions on the relevant transitions. This ensures that the Petri net only admits state progressions that satisfy the logical rules of OIDC.

    \item {\bf Initial and terminal states.} The initial marking $M_{0\phi}$ places a token in the ${<}{\sf start}{>}$ state of the protocol. Terminal markings correspond to successful completion, such as after ${<}{\sf verifyClaims}{>}$. This enables reachability analysis to check whether a candidate implementation can faithfully execute the entire flow.
\end{itemize}
Through this mapping process, each of the three OIDC authentication flows is transformed into a Petri net model that captures both the sequential ordering of protocol steps and the logical conditions imposed by the specification. This specification Petri net is then used as the reference model against which candidate program nets are validated.

\begin{example}\rm
    \autoref{fig:program-spec-petri}(B) illustrates the specification Petri net constructed from the specification (as standard Petri net input formats~\cite{pnml}) provided by the user. Each place in $\{P_1, P_2, \dots, P_n\}$ denotes a specification state and transitions denote the events that should be satisfied to enable firing. Similar to the program Petri net, specification transitions can also be associated to guard conditions using the relevant protocol variables and operators. 
\end{example}

\subsection{Validation}\label{subsec:validation}

We regard a patch as {\em correct} if the behavior of the patched program is consistent with the OpenID specification (i.e., semantically equivalent), meaning that every program trace is admitted by the specification (i.e., reachable).

Once the Petri net $\mathcal{N}_P$ is constructed for the candidate program, we validate it against the specification Petri net $\mathcal{N}_\phi$ using the \textsc{Validate} procedure described in \autoref{alg:validation}. The procedure takes as input both nets and outputs either $\top$ if the candidate program conforms to the specification, or $\bot$ together with a counterexample set $\Phi$ if the program violates the specification.  

The algorithm initializes the search space $\Phi$ with the initial markings of the two nets (line~5). In each iteration, it selects a pair of markings $(M, M_\phi)$ to explore (line~7). If the program net has reached one of its final markings while the specification net has not (line~8), the procedure terminates immediately and returns $\bot$ along with the counterexample trace $\Phi$.  

For each enabled transition $T$ in the program net (line~9), the algorithm checks whether there exists a corresponding enabled transition $T_\phi$ in the specification net (line~10). If such a transition exists and both the event label $\epsilon$ and the guard condition $\gamma$ match (line~11), then both transitions are fired simultaneously, and the resulting markings $(M', M'_\phi)$ are added to the search space (lines~12--13). Otherwise, the algorithm fires only the program transition $T$ while leaving the specification marking unchanged (lines~17--18).  

The procedure continues until the search space is exhausted. If no violating case is found, the algorithm terminates with $\top$ and an empty counterexample set $\varnothing$ (line~21), indicating that the candidate program satisfies the specification.  

\begin{algorithm}[t]
    \small
    \begin{algorithmic}[1]
        \Procedure{\textsc{Validate}}{$\net_P, \net_\phi$}
            \State {\bf input:} Petri net $\net_P$ of the candidate program $P$, and Petri net $\net_\phi$ of the specification $\phi$
            \State {\bf output:} $\top$ if valid, otherwise $\bot$ with counterexample $\Phi$
            \State {\bf assume:} $\net_p(P, T, A, M_0) \land \net_\phi(P_\phi, T_\phi, A_\phi, M_{0\phi})$
            \State $\Phi = \{\langle M_0, {M_0}_{\phi} \rangle\}$
            \While{$\Phi \neq \varnothing$}
                \State {\bf choose} $\langle M, {M}_{\phi} \rangle \in \Phi$
                \State {\bf if} $M \in {\sf final}(\net_P) \land M_\phi \notin {\sf final}(\net_\phi)$ {\bf then} {\bf return} $\bot, \Phi$
                \ForAll {$T \in {\sf enabled}(M)$}
                    \If{$\exists T_{\phi} \in {\sf enabled}(M_{\phi})$}
                        \If{$T.\epsilon = T_{\phi}.\epsilon \land T.\gamma \vDash T_{\phi}.\gamma$}
                            \State $M', M_\phi' = {\sf fire}(M, T), {\sf fire}(M_{\phi}, T_{\phi})$
                            \State $\Phi = \Phi \cup \langle M', M_{\phi}' \rangle$
                            \State {\bf continue}
                        \EndIf
                    \EndIf
                    \State $M' = {\sf fire}(M, T)$
                    \State $\Phi = \Phi \cup \langle M', M_{\phi} \rangle$
                \EndFor
            \EndWhile
            \State \textbf{return} $\top, \varnothing$
        \EndProcedure
    \end{algorithmic}
    \caption{Petri Net Based Program Validation}
    \label{alg:validation}
\end{algorithm}

\subsection{Counterexample-Guided Solution Refinement}\label{subsec:cex}

Once a candidate patch is verified against the Petri-net specification, the verification procedure \textsc{Validate} may either succeed ($\top$) or fail ($\bot$). In the case of failure, \toolname leverages the verification feedback in two complementary ways to refine the repair process:
\begin{itemize}
    \item First, the binary pass/fail outcome directly eliminates the current candidate from the search space, ensuring that invalid patches are not re-suggested in subsequent iterations.
    
    \item Second, beyond this coarse-grained signal, the Petri-net verifier also provides a counterexample trace $\Phi$ that captures the precise execution path or sequence of inputs leading to the violation of the specification. This counterexample is then reformulated as part of a structured follow-up prompt to the LLM, guiding it toward avoiding similar errors in the next round of patch synthesis.
\end{itemize}

Concretely the prompt to the LLM includes: (i) the buggy code with the previous candidate patch, (ii) the relevant specification excerpt and (iii) the counterexample trace highlighting the mismatch between the program and the specification. By incorporating the counterexample into the LLM’s context, \toolname effectively directs the model to prioritize alternative repairs that eliminate the violation. This iterative loop of {\em verification / counterexample extraction / LLM refinement} allows \toolname to converge more efficiently toward a correct and specification-compliant patch, while pruning away unproductive search directions. We provide a detailed example in \autoref{apdx:cex}.
\section{Evaluation}\label{sec:eval}

\begin{table*}[!ht]
\centering
\footnotesize
\caption{Average time for fault localization, sketch generation, enumeration, and validation in our dataset.}
\begin{tabular}{
>{\raggedright\arraybackslash}p{3.0cm}|
>{\centering\arraybackslash}p{1cm}
>{\centering\arraybackslash}p{2.6cm}
>{\centering\arraybackslash}p{2.8cm}
>{\centering\arraybackslash}p{4cm}
>{\centering\arraybackslash}p{2cm}}
\textbf{OpenID Bug Type}
& \textbf{\#Bug}
& \textbf{Fault Localization (s)}
& \textbf{Sketch Generation (s)}
& \textbf{Enumeration \& Validation (s)}
& \textbf{Total Time (s)}
\\ \midrule
Incorrect auth flow
& 2
& 2.3
& 38.1
& 294.0
& 334.4\\

Signature verification
& 7
& 5.1
& 88.7
& 583.2
& 677.0\\

\texttt{alg} validation
& 3
& 2.0		
& 25.8
& 191.5
& 219.3\\

\texttt{aud} validation
& 3
& 2.7			
& 53.2
& 249.8
& 305.7\\

\texttt{iss} validation
& 2
& 3.1			
& 47.4
& 145.3
& 195.8\\

\texttt{nonce} check
& 1
& 4.5			
& 78.2
& 265.4
& 348.1\\

Access token validation
& 2
& 4.5
& 84.5
& 363.7
& 452.7\\

CSRF protection
& 3
& 3.6
& 88.3
& 282.5
& 374.4\\

\end{tabular}
\label{tab5:repair-results}
\vspace{-1.5em}
\end{table*}

We design our evaluation scheme primarily to answer the following research questions: 
\begin{itemize}
    \item {\bf RQ1}: Is \toolname effective in fixing real-world bugs in OpenID implementations?

    \item {\bf RQ2}: How effective is our approach when compared to other LLM-based program repair methods?

    \item {\bf RQ3}: Are the generated repairs as correct as the manual patches written by the developers?
\end{itemize}

\subsection{Implementation}
We implement the technical concepts discussed above in the \toolname tool, which takes OIDC programs and their specification (in English and the corresponding Petri net representation in PNML format~\cite{hillah2010pnml}) as input and generates a fixed program that satisfies the specification. The \toolname tool, consisting of approximately 5,400 lines of Java code, uses the popular ChatGPT (GPT-3.5)~\cite{wu2023chatgpt} LLM agent for the localization of bugs and the generation of optimized repair expressions. Additionally, we utilize IBM T.J. Watson Libraries for Analysis (WALA)~\cite{wala} toolkits to generate program representations (such as call graphs and control flow graphs) when transforming program semantics into Petri net representations. Furthermore, we employ the SAT4J tool~\cite{le2010sat4j} to check the equivalence of guard conditions associated with Petri net events in both programs and specifications.

\subsection{Experimental Setup}

We describe in the following the setup of our evaluation, covering benchmarks, specifications, and environment.

\paragraph{Benchmark collection} 
To evaluate the performance of our repair approach, we compile a dataset of OpenID bugs that consists of confirmed security bugs in OpenID libraries in previous papers ~\cite{oauthlint-ase19,ccs22,ccs24}, as well as public records (\eg CVE reports) for other projects based on OpenID. Our collected dataset of OpenID bugs is representative as they are collected from the most popular OpenID libraries and cover all three authentication flows in the OpenID protocol: 1) authorization code flow (AC), 2) implicit flow (I), and 3) hybrid flow (H). Depending on the flow, the impact of each bug is either the relying party (RP), OpenID provider (OP), or both. In our dataset, each benchmark contains at least one bug that violates the OpenID specification. Along with the bugs, we further collect the patches written by human developers to fix the bugs. Specifically, our dataset contains 23 bugs from eight categories of OpenID bugs. We show detailed statistics of the benchmarks collected  in \autoref{apdx:bench-stats}.

\paragraph{Specification}
We follow OpenID Connect Core 1.0~\cite{openid-spec} to get the specification for each benchmark in our dataset. The standard specification describes the required behavior of the authentication flows that are supported by OpenID. In addition, it details the protocol parameters and how each party involved in the protocol execution should process them. Once we collect the English specification for each benchmark, we manually construct its equivalent Petri-net model, as described earlier in the paper. 

\paragraph{Alignment of programs with specifications}
For each buggy program, we adopt a semi-automatic procedure to determine the relevant Petri-net specification. We keep a library of Petri-net models for different authentication flows and security properties, and use an LLM-based retrieval-augmented generation (RAG)~\cite{rag} method to suggest the closest match. A lightweight manual confirmation then finalizes the minimal alignment between the buggy program and the specification. In this way, the Petri-net model is not global but contextually selected, and \toolname operates in a {\em modular} fashion on the function or code region associated with the bug, which can be identified by existing bug localization or vulnerability detection tools~\cite{ccs22}.

\subsection{Experimental Results}
We carried out all experiments on a machine equipped with a Quad-Core Intel Core i5 processor and 32GB of memory, operating on macOS 14.4. In total, 23 Petri nets were constructed for the benchmark programs, and 12 are constructed from the specification OpenID that covers the security properties for the three authentication flows. We delve into the details of the results of our experiments as follows.

\begin{table}[!t]
\centering
\footnotesize
\caption{Evaluation results of \toolname for OpenID bugs repair in our dataset. Column \#Fix$_{\textsf{A}}$ shows the correct fixes generated by \toolname and column \#Fix$_{\textsf{dev}}$ shows the number of generated fixes that are semantically equivalent to the manual patches written by the developers (if available).}
\begin{tabular}{
>{\raggedright\arraybackslash}p{3.5cm}|
>{\centering\arraybackslash}p{1cm}
>{\centering\arraybackslash}p{1cm}
>{\centering\arraybackslash}p{1cm}}
\textbf{OpenID Bug Type}
& \textbf{\#Bug}
& \textbf{\#Fix$_{\textsf{A}}$}
& \textbf{\#Fix$_{\textsf{dev}}$}
\\ \midrule
Incorrect auth flow
& 2
& 1
& 1 \\

Signature verification
& 7
& 4
& 2 \\

\texttt{alg} validation
& 3
& 3
& 3\\

\texttt{aud} validation
& 3
& 3
& 3 \\

\texttt{iss} validation
& 2
& 1
& 1 \\

\texttt{nonce} check
& 1
& 1
& - \\

Access token validation
& 2
& 2
& - \\

CSRF protection
& 3
& 2
& 0 \\

\midrule
\textbf{Overall}
& 23
& \textbf{17}
& 10 \\
\end{tabular}
\label{tab4:repair-results}
\vspace{-1.5em}
\end{table}

\paragraph{Effectiveness of \toolname (RQ1)}
To evaluate the effectiveness of \toolname in repairing OpenID bugs, we use it to repair 23 benchmarks in our dataset and manually check the repaired bugs to validate their correctness with respect to the corresponding specification. As shown in \autoref{tab4:repair-results}, it generated correct fixes for 17 out of 23 bugs. Among these 17 bugs, ten bugs require fixing conditional expressions and method invocations, three bugs involve fixing the return statement and four require fixing expressions used in the method's arguments. In addition, the average time for \toolname to fix a bug is 362 seconds (6 minutes), whereas the minimum and maximum times are 78 seconds and 14 minutes, respectively.

We further examined the reasons behind the six instances where \toolname failed to find a repair and identified the following root causes: (1) \toolname was unable to repair expressions that required variables that are exchanged during the dynamic registration (e.g., \icode{id_token_signed_response_alg}), which also allows passing a response as a JWT object. As extracting parameters from these objects requires complex operations such as decryption and base64 decoding, our repair algorithm could not synthesize expressions where values from these objects are required. (2) \toolname also allows the clients to obtain keys to validate ID token from an external URI set by the \icode{jwks\_uri} parameter. Our repair algorithm was unable to synthesize the API calls (e.g., HTTP calls) needed to obtain the values from these external sources.

\paragraph{Comparison with other LLM-based repair approaches (RQ2)}
Given the recent success of LLM in various program repair tasks~\cite{joshi2023repair,xia2023automated}, we further investigate how our approach performs compared against them, as follows.

1) ThinkRepair~\cite{yin2024thinkrepair}: To enhance LLM's repair capabilities, ThinkRepair~\cite{yin2024thinkrepair} utilizes a two-phase process involving a curated knowledge pool for bug fixing and a bug-fixing phase with Chain of Thoughts (CoT) prompting and few-shot learning. Specifically, given a corpus of buggy functions, ThinkRepair first builds a knowledge pool for chains of thoughts on fixing the buggy functions. To repair a bug, it then uses examples from the knowledge pool to guide LLMs in understanding and fixing bugs while adjusting prompts iteratively with test feedback to refine solutions. To compare \toolname, we utilize the  knowledge pool collected by ThinkRepair for Defects4J dataset~\cite{just2014defects4j} and use them to select examples of bugs and CoTs for similar repairs to guide the LLM to repair our OpenID benchmarks. For this selection step, we choose the contrastive-based selection as it utilizes a contrastive learning framework to further fine-tune UniXcoder for better
semantic embedding. As shown in \autoref{tab4:repair-comparison-llm}, ThinkRepair was able to repair 7 bugs (out of 23) when evaluated on our OpenID benchmarks. Since ThinkRepair utilizes bug examples from its knowledge pool to guide the repair process, we found it was most effective in repairing bugs in conditional statements. However, it performs poorly for bugs (e.g., signature validation) that require understanding the semantics of the OpenID protocol and specification. 

\begin{table}[t]
\centering
\footnotesize
\caption{Comparison between the repairs generated by \toolname (\textbf{\#Fix$_{\textsf{A}}$}), ThinkRepair (\textbf{\#Fix$_{\textsf{T}}$})~\cite{yin2024thinkrepair}, and LLM inference approach (\textbf{\#Fix$_{\textsf{L}}$}) for the OpenID bugs.}
\begin{tabular}{
>{\raggedright\arraybackslash}p{4cm}|
>{\centering\arraybackslash}p{1cm}
>{\centering\arraybackslash}p{1cm}
>{\centering\arraybackslash}p{1cm}}
\textbf{OpenID bug type}
& \textbf{\#Fix$_{\textsf{A}}$}
& \textbf{\#Fix$_{\textsf{T}}$}
& \textbf{\#Fix$_{\textsf{L}}$}
\\ \midrule
Incorrect auth flow (2)
& 1
& 0
& 0\\

Signature verification (7)
& 4
& 1
& 1\\

\texttt{alg} validation (3)
& 3
& 2
& 1\\

\texttt{aud} validation (3)
& 3
& 1
& 1\\

\texttt{iss} validation (2)
& 1
& 1
& 0\\

\texttt{nonce} check (1)
& 1
& 1
& 0\\

Access token validation (2)
& 2
& 0
& 0\\

CSRF protection (3)
& 2
& 1
& 1\\

\midrule
\textbf{Overall}
& 17
& 7
& 4\\
\end{tabular}
\label{tab4:repair-comparison-llm}
\vspace{-1.5em}
\end{table}
    
2) LLM inference: We further compare the results with LLM inference only (i.e., zero-shot completion) program repair. Following the common prompt construction approach in prior works~\cite{joshi2023repair}, we construct prompts to repair the bugs in each of our benchmarks. Specifically, we provide the LLM agent with the buggy code from the benchmarks in our dataset and a natural language specification, asking it to fix the corresponding bugs. We submit an LLM query $Q = \{q_i, q_{code}, q_{spec}\}$ where $q_i$ is a static repair instruction for the LLM agent, $q_{code}$ is the buggy code and $q_{spec}$ is the OpenID specification for the corresponding bug. As shown in \autoref{tab4:repair-comparison-llm}, LLM inference approach was able to generate repairs for only four bugs. We observe that LLM can fix commonly observed programming bugs~\cite{saha2017elixir,xuan2016nopol} such as missing clauses for null checking. However, in this work, we focus on the logical bugs relevant to the incorrect implementation of OpenID protocol, for which LLM agents struggle to generate a valid repair. 

\paragraph{Quality of the generated repairs (RQ3).}
To answer our second research question, we manually investigate the generated patches and compare them against the patches written by the developers of the libraries we include in our benchmarks. These manually written patches are based on the bugs reported in prior works~\cite{ccs22,ccs24} and their respective patches based on the GitHub commit history. However, for three bugs (from the categories of `nonce check' and `access token validation'), we were not able to obtain the manual patches as the developers had not published any patches by the time of our experiments. In total, we collected the 20 manually written patches corresponding to the bugs in our dataset. Upon our manual investigation, we found that 10 out of the 15 generated patches are semantically equivalent to the patches written by human developers (\autoref{tab4:repair-results}). The other six patches are not exactly semantically equivalent just because our patches used hard-coded values (e.g., OpenID configured constants) whereas developers obtained the values from dynamic API calls or from sessions. These results show that \toolname can generate high-quality patches.

\subsection{Threats to Validity} 

This section outlines possible limitations of our study and their implications for the results.

\paragraph{Benchmark creation}
Our benchmark doesn't include all the publicly reported bugs in OpenID implementations. As many reported bugs include the OpenID implementations that are not open-sourced, creating such a comprehensive dataset is quite challenging. However, we still tried our best efforts to select the most representative bugs and cover a wide variety of bugs in our benchmarks. Furthermore, since \toolname is designed to cover all the flows in OpenID protocol and incorporate domain knowledge of the entire OpenID specification, we believe \toolname can generalize well to new OpenID bugs. 

\paragraph{Petri net creation}
The Petri net models were manually constructed to capture the program’s expected behavior according to the specification. On average, constructing the model for each of the three OpenID authentication flows took about 3.5 hours and required expertise in formal modeling and the OIDC specification. Since this is a one-time effort per flow, the resulting models can be reused across all benchmarks following the standard. Although effective in practice, prior work~\cite{doc2spec} shows that formal specifications can be automatically extracted from natural language documentation, suggesting that future research could further reduce or even automate this effort using LLM-based techniques.
\section{Related work}\label{sec:related}

In this section, we review related lines of work and highlight how our approach differs from and complements them.

\subsection{Search-Based Repair}

GenProg~\cite{le2011genprog}, a pioneering work in the area of automated program repair utilizes genetic programming to explore a search space of potential repairs generated by reusing code snippets from within the program. PAR~\cite{kim2013automatic} utilizes GenProg’s search approach with a set of repair templates manually derived from human-written patches. Le et al.~\cite{le2016history} employs an extensive collection of templates sourced from GenProg, PAR, and mutation testing to generate a diverse array of potential repairs. These repairs are subsequently organized and pruned based on the frequency of similar (human-written) patches. Later, ACS~\cite{xiong2017precise} introduces a technique for accurate condition synthesis by instantiating variables within commonly occurring predicates across a designated code corpus, employing various heuristics to prioritize and select the most suitable variables. In contrast, RSRepair~\cite{weimer2013leveraging} employs a random search approach, while AE~\cite{qi2014strength} leverages deterministic search, enhanced by analytical techniques to eliminate redundant patches and optimize the search process.

Different from previous work, our proposed approach tackles the repair problem for complex logical bugs in OpenID Connect protocol that requires a richer domain-specific repair expression to fix bugs. Moreover, existing repair techniques rely on comprehensive test cases or human-written manual patches, which are not often available for complex protocols like OpenID Connect.

\subsection{Repair Using Constraint Solving}

MintHint~\cite{kaleeswaran2014minthint}, and NOPOL~\cite{xuan2016nopol} employ symbolic execution to construct oracle-like representations and then use program synthesis to generate repairs. DirectFix~\cite{mechtaev2015directfix} targets concise fixes via partial maximum satisfiability with SMT formulas, while Angelix~\cite{mechtaev2016angelix} improves scalability with a lightweight constraint mechanism. S3~\cite{le2017s3} further augments these semantics-based approaches with ranking criteria from execution traces, and CPR~\cite{shariffdeen2021concolic} leverages concolic path exploration to prune overfitting patches. More recently, SymlogRepair~\cite{symlogrepair} combines repair with Datalog-defined static analysis, showing that constraint solving can capture richer program properties.

These methods, however, require translating constraints into SAT/SMT, which risks incompleteness~\cite{mechtaev2016angelix,le2017s3} due to complex semantics and external libraries common in API-based protocols like OpenID Connect. They also mainly reason about boolean or integer conditions, limiting their ability to handle bugs involving complex APIs such as cryptographic functions. Furthermore, symbolic execution engines (\eg KLEE~\cite{mechtaev2016angelix}) often fail to extract constraints at scale. In contrast, our approach operates directly at the AST level, avoiding heavy translation and enabling the repair of expressions, diverse variable types, function invocations, and APIs, while remaining applicable to complex data structures as long as they can be executed.

\subsection{AI for Program Repair}

In recent years, AI-based methods, especially Convolutional Neural Networks (CNNs), Neural Machine Translation (NMT), LLMs, and their combination have shown success in program repair. Lutellier et al.~\cite{lutellier2020coconut} proposes CoCoNuT which leverages ensemble learning, combining CNN and NMT to generate patches. DLFix~\cite{li2020dlfix} adopts a two-tier approach, where the first layer learns the context of bug fixes, and the second layer generates the corresponding patch. Notably, CURE~\cite{jiang2021cure} has recently achieved state-of-the-art results on the Defects4J~\cite{just2014defects4j} and QuixBugs~\cite{ye2021quickbugs} datasets, outperforming NMT-based APR techniques by utilizing a pre-trained programming language model, code-aware search, and sub-word tokenization. Elixir~\cite{saha2017elixir} uses machine learning to rank potential repair candidates to reduce the search space for object-oriented programs. InferFix~\cite{jin2023inferfix} combines LLM and static analyzer to fix critical security and performance bugs. AI-based approaches rely on extensive datasets of patched programs, tailored to specific bugs. However, the scarcity of reported and patched bugs in the context of the OpenID Connect makes it impractical to develop a robust, generalized learning model for repairing such bugs.
\section{Conclusion}\label{sec:conclusion}

OpenID Connect has significantly improved online authentication, but its complexity has led to critical security vulnerabilities causing substantial financial and data breaches. To address this, our proposed tool, \toolname, leverages large language models to automate bug detection and patch generation, ensuring accurate and reliable fixes through a novel Petri-net-based model checker. Our evaluation of a dataset of OpenID bugs demonstrates that \toolname successfully generates correct patches for 17 out of 23 bugs (74\%), with a high proportion of patches semantically equivalent to developer-written fixes. 
\section*{Acknowledgments}

This work is supported in part by Google Faculty Research Award, Ethereum Foundation Academic Award, NSF 1908494, and DARPA N66001-22-2-4037.

\bibliographystyle{IEEEtranS}
\bibliography{ref}

\clearpage
\appendices
\crefalias{section}{appendix}
\crefname{appendix}{Appendix}{Appendices}
\Crefname{appendix}{Appendix}{Appendices}

\begin{table*}[!ht]
\centering
\footnotesize
\caption{Statistics of our dataset of OpenID bugs in Java that are reported in prior works~\cite{oauthlint-ase19,ccs22} of OpenID and OAuth security analysis. These bugs may impact participating entities such as the relying party (RP) and OIDC provider (OP) while interacting using the standard flows of Authorization Code (AC), Implicit (I), and Hybrid flows (H). }
\begin{tabular}{
>{\raggedright\arraybackslash}p{4cm}|
>{\centering\arraybackslash}p{1.5cm}
>{\centering\arraybackslash}p{1.5cm}
>{\centering\arraybackslash}p{2cm}
>{\centering\arraybackslash}p{2.5cm}
>{\centering\arraybackslash}p{2.8cm}}
\multicolumn{1}{l}{\textbf{Bug Type}}
& \textbf{\#Bugs}
& \textbf{\#LOC}
& \textbf{Auth Flows}
& \textbf{Impacted Parties}
& \textbf{Spec. Reference}
\\ \midrule
Incorrect auth flow
& 2
& 342
& AC, I, H 
& RP, OP
& \S3.1, \S3.2, \S3.3
\\

Signature verification
& 7
& 420
&  I, H
& RP, OP
& \S3.1.3.6--7 
\\

\texttt{alg} validation
& 3
& 235
&  I, H
& RP, OP
& \S3.1.3.7  
\\

\texttt{aud} validation
& 3
& 230
& AC, I, H
& RP
& \S3.1.3.7  
\\

\texttt{iss} validation
& 2
& 148
& AC, I, H
& RP
& \S3.1.3.7  
\\

\texttt{nonce} check
& 1
& 198
& I, H
& RP, OP
& \S3.1.3.7   
\\

Access token validation
& 2
& 348
& AC, I, H
& RP
& \S3.1.3.7, \S3.2.2.9
\\

CSRF protection
& 3
& 186
& AC, I
& RP, OP
& \S3.1.2.1
\\


\end{tabular}
\label{tab:bug-stats}
\vspace{-1.5em}
\end{table*}

\begin{figure}[h]
    \begin{minted}[xleftmargin=5pt,numbersep=1pt,fontsize=\scriptsize,linenos,breaklines]{md}
### Tasks
Consider the given buggy program sketch and compare the specification description to understand the intended behavior.  
Now, analyze the failure traces, especially the events and guard conditions in failure points, to identify why the candidate patch does not meet the specification and propose a new candidate patch that corrects the identified issues. Preserve correct fixes already present in the candidate patch and output only the new candidate patch, nothing else.
### Buggy program/sketch
{buggy_program_sketch}
### Specification
{spec_description}
### Candidate patch
{candidate_patch}
### Counterexample trace (from verification)
{
  "candidate_id": "b1-c25",
  "verification_status": "failed",
  "failure_point": {
    "transition": "T3",
    "program_state": {
      "current_place": "B3",
      "next_place": "B6",
      "marking": [1, 1, 1, 1, 0, 0, 0]
    },
    "event": {
      "specification": {
        "name": "validatePlain",
        "parameters": ["id_token"]
      },
      "program": {
        "name": "validatePlain",
        "parameters": ["id_token"]
      }
    },
    "guard_conditions": {
      "specification": {
        "conditions": [
          "alg == none",
          "allowedUnsigned == true"
        ]
      },
      "program": {
        "conditions": [
          "alg == null"
        ]
      }
    }
  }
}
    \end{minted}
    \caption{An example prompt that encodes counterexample information provided to LLM.}
    \label{fig:cex-prompt}
    \vspace{-1.5em}
\end{figure}

\section{Counterexample-Guided Solution Refinement}\label{apdx:cex}

\autoref{fig:cex-prompt} shows an example of the inputs used to refine candidate patches. The example employs a structured prompt that provides the LLM with the necessary components to adjust patches based on prior verification failures. The buggy program sketch and the natural language specification describe the intended behavior, while the candidate patch corresponds to the most recent attempt that failed verification. The counterexample trace records the failing transition along with the relevant program state, event alignment, and guard condition mismatch that caused the violation. This information gives the LLM targeted feedback on why the candidate patch did not satisfy the specification, such as the absence of the required guard condition \icode{allowedUnsigned == true}. As a result, the LLM is guided to generate subsequent patches that differ from the rejected candidate while incorporating the specific failure information.

\section{Benchmark Statistics}\label{apdx:bench-stats}

\autoref{tab:bug-stats} summarizes the statistics of our benchmark of OpenID bugs in Java, as collected from prior security analyses. The dataset covers eight representative categories of vulnerabilities, including incorrect authentication flows, signature and token validation errors, and missing CSRF protection. In total, the benchmark contains 23 bugs, with each bug localized to a code region of moderate size (148–420 lines of code). The bugs span all three standard OpenID Connect authentication flows—Authorization Code (AC), Implicit (I), and Hybrid (H)—and affect both major protocol participants, namely the relying party (RP) and the OpenID provider (OP). Each bug can be traced to specific sections of the OIDC specification, allowing us to align the faulty implementation with its corresponding normative requirement.

\end{document}